\documentclass[twoside]{dis09}
\usepackage[dvips]{graphicx,epsfig,color}
\usepackage{wrapfig,rotating}
\usepackage{amssymb,amsmath,array}

\begin{document}
\title{Limitations on Dispersion Relations for Generalized Parton Distributions}

\author{Gary R. Goldstein$^1$ and Simonetta Liuti$^2$
%
\thanks{Partially supported by U.S. Department of Energy, DE-FG02-01ER4120(S.L) \& DE-FG02-92ER40702(G.R.G.)}
%
\vspace{.3cm}\\
%
1- Department of Physics and Astronomy \\
Tufts University - Medford, MA 02155, USA%
\vspace{.1cm}\\
2- Department of Physics \\
University of Virginia - Charlottesville, VA 22901, USA.\\
}

\maketitle

\begin{abstract}
Deeply Virtual Compton Scattering and electroproduction of mesons involve amplitudes that are analytic in energy. Analyticity enables Dispersion Relations (DR's) for such amplitudes, relating real and imaginary parts. Lately it has been suggested that DR's be applied to the integrated Generalized Parton Distributions that embody the spin-dependent soft, but factorizable part of the scattering. However, at non-zero momentum transfer, DRs require integration over unphysical regions of the variables. We show that the relevant unphysical region of the non-forward DRs is considerable, vitiating efforts to avoid the actual measurement of the real parts more directly.
\end{abstract}

\begin{wrapfigure}{r}{0.5\columnwidth}
\centerline{\includegraphics[width=0.45\columnwidth]{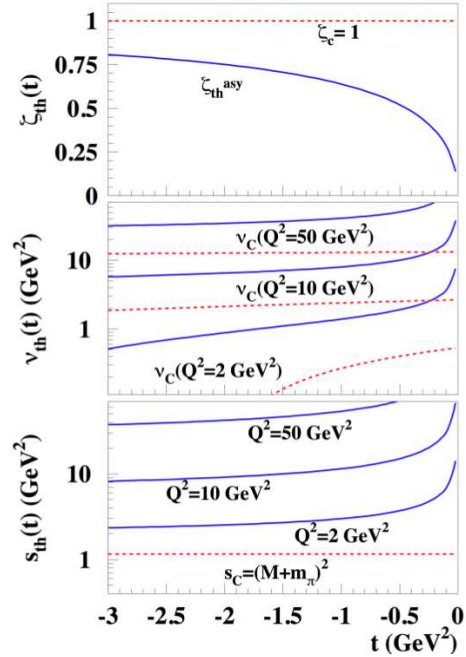}}
\caption{Thresholds for the variables $\zeta$, $\nu$ and $s$ plotted vs. $t$.  The dashed lines are the continuum thresholds whereas
the full lines represent the physical thresholds (see text).}\label{fig1}
\end{wrapfigure}

A number of studies have recently advocated using Dispersion Relations (DRs) both to facilitate the extraction from deeply virtual exclusive experiments, such as Deeply Virtual Compton Scattering (DVCS), of the soft matrix elements  for hard exclusive processes, the Generalized Parton Distributions (GPDs), and to determine  their model parameters \cite{AniTer,DieIva,BroEst}. In this presentation we will show that there are important limitations to the use of DRs for processes described by GPDs~\cite{url}.

DRs  have a long history in hadronic physics. For a general exclusive, two body hadronic reaction, invariant amplitudes can be written in terms of energy and angle variables, such as the Mandelstam variables $s$  and $t$,  or $\nu = (s-u)/4M$ and $t$. 
When the energy variable is continued into the complex plane, the amplitudes become holomorphic functions, i.e. analytic functions over regions of the complex plane. Unitarity of the amplitudes determines the physical intermediate states that, in turn determine branch cuts in the complex energy plane. Each physical state has a kinematic threshold that fixes the branch point. 

DRs were derived for $inclusive$ Deep Inelastic Scattering (DIS)  as well, when viewed as forward virtual Compton scattering \cite{jaffe}.    
DIS can be considered as a special case of elastic scattering where unitarity relates the imaginary part of the forward amplitude to the total cross section, the inclusive sum over all physical final states allowed by the energy. 
As discussed thus far, DRs do not  necessarily include the partonic structure 
of the target. Partonic degrees of freedom are integrated over.  In fact, all remaining kinematical variables, including $x_{Bj}$, 
can be considered to be fixed by the kinematic conditions ``external" to any partonic loop or QCD elaboration.  
A connection with the partonic structure,  through the Operator Product Expansion, therefore QCD, can be established {\it e.g}. by following the derivation in Ref.~\cite{jaffe}, 
where the important assumption is made that the support for both the integrals defining the Mellin moments of the 
operators and the final amplitude is $x_{Bj}   \in [-1,1]$, in the asymptotic limit, $Q^2 \rightarrow \infty$. 
More specifically,  
two steps are taken in establishing DRs for DIS: {\it i)} the identification of the physical threshold for the
scattering process, $\nu_{th}$, with the continuum threshold, $\nu_C$ for $s=(M+m_\pi)^2$, or $x_C=1$,  appearing in the integral definition of the scattering amplitude; {\it ii)} the identification of $x_{Bj}$ with the partonic variable present in the factorized amplitude.  

In this presentation~\cite{url} we argue that these assumptions cannot be carried straightforwardly to the off-forward case described {\it e.g.} in DVCS. In fact, 
one has  a mismatch between the supports for the scattering amplitude and for the corresponding DR, namely $\nu_{C} \neq \nu_{th}$. This is
a straightforward consequence of $t$-dependent physical thresholds, not present in the DIS forward/elastic case, that are long known to hinder the useful and practical applications of DRs. The mismatch exists for both $\nu$ and $\zeta$. This point does not touch upon the partonic aspects of the process.  However,
in the factorized form of DVCS,  described by a  handbag picture, 
$x_{Bj}$ in  {\it ii)}  is replaced by two longitudinal fractions, $X$ and $\zeta$,  where $X \equiv (kq)/(Pq)$ 
and the external variable, the skewness, 
$\zeta = (q \Delta )/(q  P)  \approx Q^2/(2M \nu_{Lab})$,
$\Delta$ being the momentum transfer for the two body scattering process, $\Delta^2=t$
(we will use either the set $(X,\zeta,t)$ or the alternative variables, $(x=\frac{X-\zeta/2}{1-\zeta/2}, \xi=\frac{\zeta}{2-\zeta},t)$ throughout the paper- see Ref.\cite{BelRad,Die_rev} for reviews on DVCS). 
The expression for the DVCS amplitude in QCD factorization is
\begin{equation}
\label{DVCS}
T^{\mu\nu}(\nu,Q^2,t)=\frac{1}{2}g^{\mu\nu}{\bar u}(p^\prime){\hat n}u(p)
\sum_{flavors} e_f^2 {\cal H}_f^{(\pm)} (\xi,t),
\end{equation}
where the analog of the Compton Form Factor (CFF)  is
\begin{equation}
\label{direct}
{\cal H}_f ^{(\pm)}(\xi,t)=\int\limits_{-1}^{+1}dx \frac{H_f ^{(\pm)}(x,\xi,t)}{x-\xi+i\epsilon}.
\end{equation}
Crossing symmetry is implemented by
$H_f^{(\pm)}(x,\xi,t)=H_f(x,\xi,t) \mp H_f(-x,\xi,t)$,
recalling that for PDFs, $q(-x)=-{\bar q}(x)$ relates negative $x$ to positive $x$ antiquark probability.

It follows straightforwardly from Eq.(\ref{direct}) that  ${\rm Im} \, {\cal H}(\xi,t) = -\pi H(\xi,\xi,t)$. To relate this to the discontinuity across the physical branch cut of a holomorphic function, unitarity is invoked through the insertion of a complete set of intermediate states. 
\begin{equation}
{\rm Im} \mathcal{H}(\zeta,t) =  2\pi \int dX \left[ \delta(X-\zeta) + \delta(X) \right] 
 \sum_N  
 \langle P^\prime \mid {\bar \psi}^+(k^\prime)  \mid N  \rangle \langle N \mid \psi^+(k) \mid P \rangle 
\delta(XP^+ + p_N^+ -P^+)  
\label{sumN}
\end{equation}
The resulting analytic structure allows the DR to be written,
\begin{equation}
{\rm Re} \, {\cal H}^{(\pm)}(\xi,t)   =   \frac{1}{\pi} \left[ P.V. \int\limits_{0}^{\xi_{th}}dx  
\frac{H^{(\pm)} (x,x,t)}{x-\xi}  \right.
+  \left.  \int\limits_{\xi_{th}}^{+1}dx  
\frac{H_{unphys}^{(\pm)} (x,x,t)}{x-\xi}  + (\xi \rightarrow -\xi) \right],
\label{xDR}
\end{equation}
where $unphys$ emphasizes that 
the integration should be over the whole range. Because the integration variable is now interpreted as the 
skewness, ``external" to the quark loop, a threshold 
mismatch ensues due to the inelasticity of the two body process for non-zero $t$. In fact 
$\zeta_{th} = [-t + (t^2 -4M^2t)^{1/2}]/2M$ for $Q^2>>t$, the physical threshold for the two body, $\gamma^* p \rightarrow \gamma p^\prime$ scattering process, originates from the limiting values for the angles defining the invariant $t=(q-q^\prime)^2$. One obtains in the limit $Q^2>>t$, $t_{min} = Q^4/4s - (q^{CM} - q^{\prime \, CM})^2= - M^2 \zeta^2/(1-\zeta)$. 
Notice that for DIS, the physical and continuum thresholds coincide because the final photon has the same $Q^2$ as the initial one, $t_{min}=0$ and $\zeta_{th}=x_{th}=1=\zeta_C$. 
\begin{wrapfigure}{r}{0.5\columnwidth}
\centerline{\includegraphics[width=0.45\columnwidth]{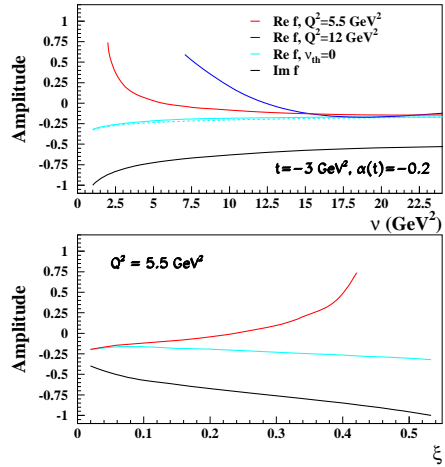}}
\caption{Threshold induced violations of dispersion relations for the Regge model; the gap between 
the calculated real parts using the physical threshold at $t=0.3 GeV^2$ and different values of $Q^2$, and the analytic continuation labeled as $\nu_{th}=0$.}
\label{fig2}
\end{wrapfigure}

In DVCS the region $x \in [\xi_{th},1]$ is unphysical and the second term in Eq.(\ref{xDR}) cannot be obtained from experiment. The physical meaning of this discrepancy is illustrated in Fig.\ref{fig1} where both  the
continuum and physical thresholds for several variable describing DVCS, $s$, $\nu$ and $\zeta$, are plotted as a function of $t$. 
For $s$, as $Q^2$ increases, only higher and higher invariant mass states are sampled. 
Although the mismatch between physical and continuum thresholds addresses  the issue of the physical 
interpretation of GPDs, it was a well known problem for two body scattering processes \cite{Lehmann} where it
was dealt with by 
either constructing  models for the analytic continuation, or developing some other prescription. 
We will show the consequences of introducing a jet mass in the factorized picture \cite{Collins}.

By analogy with the hadronic DR, it is assumed that there is a {\it physical branch cut} from -1 to +1 on the real $\xi$ axis and no other poles or cuts. For this interpretation however, the intermediate states, the $\hat{s}$-channel cuts, have to be determined, given non-zero $t$ and $Q^2$. But for these kinematic constraints the support is limited, as  Eq.(~\ref{xDR}) indicates. 
A separate consideration, is that 
intermediate states carry bare color, so there is no analog of unitarity for factorized DVCS. In DIS this distinction is irrelevant, but here 
however, the absence of intermediate hadronic states, means that the GPD cannot have the proper physical branch cuts. 
Fig.~\ref{fig1} shows that the gap remains even at high $Q^2$.

The suggestion \cite{DieIva}  
that experimental analyses provide information \emph{only on the kinematical domain on a ridge at $x=\xi$ and fixed $t$ and $Q^2$}
therefore depends on whether one can disregard or treat otherwise the unphysical term in Eq.(\ref{xDR}) 
%
It is this point about the sufficiency of the ``ridge''that we are examining with care,
by assuming
DRs are satisfied in various model GPDs. 
\begin{wrapfigure}{r}{0.5\columnwidth}
\centerline{\includegraphics[width=0.45\columnwidth]{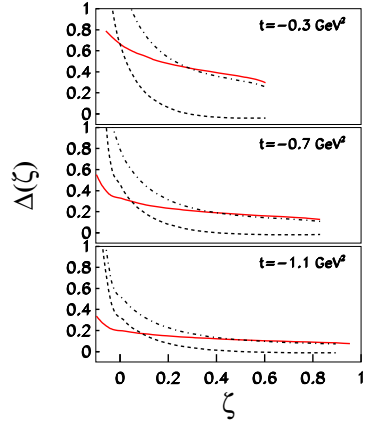}}
\caption{Simple covariant diquark model: Real part, directly and from dispersion relations and their difference.}\label{fig3}
\end{wrapfigure}
\begin{wrapfigure}{r}{0.5\columnwidth}
\centerline{\includegraphics[width=0.45\columnwidth]{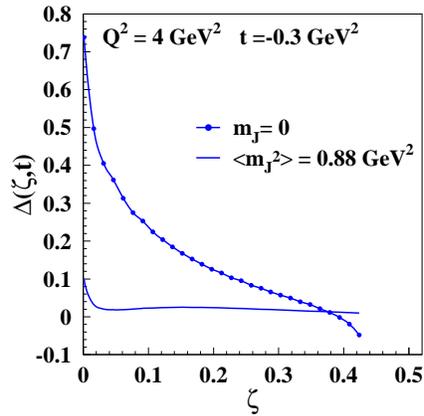}}
\caption{Difference between the dispersion relation and direct calculation in a scalar quark-diquark model
including the hard jet hadronic mass as from Eq.(\ref{jetmass2}).} 
\label{fig4} 
\end{wrapfigure}

To illustrate  these crucial questions we consider two examples of models that 
should satisfy DRs,  an asymptotic  Regge model and a covariant spectator model.
The Regge pole model  contributes to the scattering amplitude amplitude $T(\nu,t,Q^2)$ for a single Regge trajectory $\alpha(t)$ in the simple form
\begin{equation}
T^R(\nu,Q^2,t)=\beta(t,Q^2)(1-e^{i\pi\alpha(t)})\left(\frac{\nu}{\nu_0}\right)^{\alpha(t)}.
\end{equation}
So the DR should be satisfied
providing that the integral converges. For a low lying trajectory or large enough $t$ so that $\alpha(t)<0$, this will converge. But the DR relation is exact only for $\nu_{th}=0$. The actual threshold for $Q^2=0$ is at 
$-t/4M$ and further for non-zero $Q^2$. So the DR is satisfied asymptotically, for $\nu>>\nu_{th}$. This is illustrated for several cases in Fig~\ref{fig2}  where the real and imaginary parts are plotted against $\nu$ and against $\zeta$. The directly calculated real part and the dispersion relation result for the real part in this unsubtracted dispersion relation are quite separated for low $\nu$ or high $\zeta$. 

We next consider a quark diquark model with 
spinless partons (for simplicity, as in Ref.~\cite{BroEst}. 
Because this is a covariant model it  satisfies the polynomiality condition,
so analyticity requirements apply \cite{AniTer}. 
The subtraction $\Delta(\zeta,t)$, the difference between the evaluation of Eq.(\ref{direct}) and Eq.(\ref{xDR}), 
is presented in Fig.\ref{fig3}, which clearly displays non negligible $\zeta$ and $t$ variations of   
$\Delta(\zeta,t)$, thus demonstrating that $\Delta(\zeta,t)$ cannot be identified with a dispersion subtraction constant.  
In this case, since all $x,\zeta$ and $t$ dependences are part of the model in a non-trivial way, the threshold $\zeta_{max}$ is necessarily the  physical one. 
Given that our subtraction ``constant" $\Delta(\zeta,t)$ is actually a function of $\zeta$, due to the threshold dependence, we cannot see a direct relation to the so-called D-term~\cite{AniTer}.

We have not addressed the nature of the states yet. 
%
For GPDs  some kind of a duality model needs to be introduced that makes the colored quark and remnant jets
look like hadrons (see recent study on this subject  \cite{Teryaev}), 
in addition to providing a prescription for analytically continuing to the appropriate threshold.
%
The prescription we suggest  as an alternative to analytic continuation aims at reducing the kinematical threshold mismatch by 
replacing the variables used in Fig.\ref{fig1} and Fig.\ref{fig2} with variables including a mass, $m_J$, for the hard partonic jet.
%
Although considering jets with mass is not equivalent to hadronization, it might get us closer to 
what a hadronic intermediate state is. 
Following \cite{Collins,AccQiu}  we replace the hard propagators for the struck quark in the hard part of the handbag with a variable jet mass
The dispersion relation becomes
\begin{equation}
 {\rm Re} {\mathcal H} = P \int dX \int dm_J^2 \rho(m_J^2) \frac{H(X,\left(1+\frac{m_J^2}{Q^2} \right)X,t) }{\zeta-X},
\label{jetmass2}
\end{equation}  
where $\rho(m_J^2)$ is a jet mass distribution. The results shown in Fig.\ref{fig4} demonstrate that the 
gap obtained as a result of having two different thresholds in the massless calculation (Fig.\ref{fig3}) is softened, due to the new set of 
variables that better account for the correct range of integration over the partons' virtuality and transverse momentum. 
 
In conclusion, we have  shown the limitations of applying DRs to deeply virtual exclusive processes, and have given 
insight  into the partonic nature of GPDs by examing the role of variables external and internal, respectively, to the quark loop that appears in the leading order factorization formulation.  In particular, we show that it could lead to misleading results to base global parametrizations on DRs as recently done in \cite{KumMul}. To pin down GPDs we advocate comprehensive  measurements of both the  
real and imaginary components through various asymmetries and cross section components in a wide range 
of all kinematical variables, $\zeta$, $t$ and $Q^2$.
  
{\bf Acknowledgments}
We thank the organizers of DIS09 for the opportunity to present this work and to John Ralston for useful comments.



\begin{footnotesize}




\begin{thebibliography}{99}
\bibitem{url} Slides: \\ 
\verb$http://indico.cern.ch/contributionDisplay.py?contribId=194&sessionId=25&confId=53294$
\bibitem{AniTer} I.V.~Anikin and O.V.~Teryaev, Phys.\ Rev.\ {\bf D76} 056007 (2007). 
\bibitem{DieIva} M.~Diehl and D.Yu~Ivanov, Eur.\ Phys.\ Jour.\ {\bf C52} 919 (2007).
\bibitem{BroEst} S.~J.~Brodsky and F.~J.~Llanes-Estrada,
  Eur.\ Phys.\ Jour.\  {\bf C46} 751 (2006)
 \bibitem{jaffe} R.L. Jaffe, Nucl.\ Phys.\  {\bf B229} 205 (1983).
\bibitem{BelRad} A.~V.~Belitsky and A.~V.~Radyushkin,
  Phys.\ Rept.\  {\bf 418} 1 (2005)
\bibitem{Die_rev} M.~Diehl,
Phys.\ Rept.\  {\bf 388} 41 (2003).  
\bibitem{Lehmann} H.~Lehmann, Nuovo\ Cim.{\bf 10} 579 (1958).
\bibitem{Collins} J.C.~Collins, T.C.~Rogers and A.M.~Stasto, Phys.\ Rev.\ {\bf D77} 085009 (2008).
\bibitem{AccQiu} A.~Accardi and J.~W.~Qiu,
  JHEP {\bf 0807} 090 (2008).
\bibitem{Teryaev}I.~V.~Anikin, I.~O.~Cherednikov, N.~G.~Stefanis and O.~V.~Teryaev,
  arXiv:hep-ph/0806.4551 (2008). 
\bibitem{KumMul} K.~Kumericki and D.~Mueller,
  arXiv:hep-ph/0904.0458 (2009).
\end{thebibliography}
%

\end{footnotesize}


\end{document}